\documentclass[final,3p,times,twocolumn]{elsarticle}

\usepackage{graphicx}

\usepackage{amssymb}

\usepackage{epsfig}
\usepackage{amsmath}
\usepackage[active]{srcltx}

\journal{Physics Letters B}

\newcommand{\be}{\begin{equation}}
\newcommand{\ee}[1]{\label{#1} \end{equation}}
\newcommand{\ba}{\begin{eqnarray}}
\newcommand{\ea}[1]{\label{#1} \end{eqnarray}}
\newcommand{\nl}{\nonumber \\}

\begin{document}

\begin{frontmatter}


 \ead{karoly.uermoessy@cern.ch}

\title{Microcanonical Jet-fragmentation in proton-proton collisions at LHC Energy}

\author{K. Urmossy, G. G. Barnaf\"oldi and T.~S.~Bir\'o}


\address{Institute for Particle and Nuclear Physics, Wigner RCP of the HAS\nl29-33 Konkoly--Thege Mikl\'os Str., H-1121 Budapest, Hungary}


\begin{abstract}
In this paper, we show that the distribution of the longitudinal momentum fraction of charged hadrons $dN/dz$ inside jets stemming from proton-proton collisions at $\sqrt{s}$ = 7 TeV center of mass energy can be described by a statistical jet-fragmentation model. This model combines microcanonical statistics and super-statistics induced by multiplicity fluctuations. The resulting scale dependence of the parameters of the model turns out to be similar to what was observed in electron-positron annihilations in Ref.~\cite{bib:ukee}.

\end{abstract}

\begin{keyword}
microcanonical \sep fragmentation \sep proton-proton collisions \sep jets \sep superstatistics \sep Tsallis-statistics


\end{keyword}

\end{frontmatter}



\date{\today}

\section{Introduction}

When calculating the spectrum of hadrons produced in high-energy proton-proton ($pp$) collisions using perturbative quantum chromodynamics (pQCD) improved parton model calculations \cite{bib:pQCD}, hadron production is described by fragmentation functions \cite{bib:DzAKK,bib:DzKretzer}. Though the evolution of these fragmentation functions with the scale $Q^2$ can be understood within the framework of pQCD \cite{bib:DGLAPgl,bib:DGLAPap,bib:DGLAPd}, their actual form at a given scale $Q^2 = s_0$ can not be deduced from QCD. In Refs.~\cite{bib:ukee,bib:ukeeL,bib:ukeeK}, it has been shown that fragmentation functions of charged hadrons, $\pi$-s, $K$-s and $\Lambda$-s produced in electron-positron ($e^+e^-$) collisions can be described by a simple statistical physical model. This model treats hadrons formed in a jet as a microcanonical ensemble and obtains fragmentation functions via smearing the single hadron distribution in a jet over the distribution of charged hadron multiplicity measured in $e^+e^-$ annihilations. The hadron multiplicity distribution is an imput in this model and its derivation is out of the scope of this paper. Such calculations in the microcanonical ensemble have been performed by other groups to describe hadron multiplicity distributions from $e^+e^-$ to nucleus-nucleus ($AA$) collisions \cite{bib:micro1}~-~\cite{bib:micro1.2}.

The hadron multiplicity distribution used in Ref.~\cite{bib:ukee}, namely the Euler-Gamma distribution, belongs to the family of the Koba\,--\,Nielsen\,--\,Olesen (KNO) type distributions \cite{bib:kno1,bib:kno2} that describe measurements in $e^+e^-$ collisions well \cite{bib:kno3,bib:kno4,bib:kno5}.

If hadrons created in a single event are distributed according to the Boltzmann\,--\,Gibbs distribution, and the multiplicity or the temperature parameter of the distribution fluctuates according to the Euler-Gamma function, then the average hadron spectrum will be the Tsallis\,--\,Pareto (or q-canonical) distribution \cite{bib:beck2}~-~\cite{bib:wilk2}. Similarly, if in a single event hadrons have microcanonical distribution, and multiplicity fluctuates according to the Euler-Gamma distribution, then the resulting average hadron spectrum will be a microcanonical generalisation of the Tsallis\,--\,Pareto (or q-microcanonical) distribution \cite{bib:ukee,bib:ukeeL}. It is interesting, that the q-microcanonical distribution can be obtained within the framework of non-additive thermodynamics too \cite{bib:ZerothLaw}.

In this paper, we point out that the fragmentation functions measured in $pp$ collisions at $\sqrt s$ = 7 TeV collision energy \cite{bib:atlasFFpp7TeV,bib:atlasFFpp7TeV_low} can be described by the q-microcanonical distribution too. Furthermore, the parameters of the distribution show similar scale dependence as seen in Refs.~\cite{bib:HotQuarks2010,bib:Gribov80}.

The structure of this paper is as follows: in Section~\ref{mFF}, we provide a description of the microcanonical fragmentation model \cite{bib:ukee}. Section~\ref{sec:res} contains the fits of the model to fragmentation functions and multiplicity distributions measured in $pp$ collision at $\sqrt{s}$ = 7 TeV center of mass energy at the LHC \cite{bib:atlasFFpp7TeV,bib:atlasFFpp7TeV_low}. Finally, the conclusion is presented in Section~\ref{sec:con}.

\section{Microcanonical \textit{Jet} Fragmentation}
\label{mFF}

If the process of the creation of hadrons $h_1,\,\dots\,,\,h_N$ by the leading parton $p_{L}$ of a jet with multiplicity $N$ is such that the corresponding cross-section is simply proportional to the phase space available for the hadrons, restricted only by the energy conservation,

\ba
d\sigma^{h_1,\,\dots,\,h_N} &=& |\mathcal{M}|^2\,\delta^{(4)}\left(\sum\limits_j p^\mu_{h_j} - P^\mu_{p_L} \right)\,d\Omega\nl
&\propto& \delta\left(\sum\limits_j \epsilon_{h_j} - E_{jet} \right)\,d\Omega\;,
\ea{mic1}
then the hadrons created in the fragmentation process form a microcanonical ensemble. In Eq.~(\ref{mic1}), $\Omega$ is the phase space of the created hadrons, $p^\mu_{h_j}$ is the four-momentum, $\epsilon_{h_j}$ is the energy of the hadron $h_j$, $P^\mu_{p_L}$ is the four-momentum of the leading parton $p_{L}$, $E_{jet} = P^0_{p_L}$ is the energy of the jet, and $\mathcal{M}$ is the matrix amplitude describing the process. This way, neglecting hadron masses, the energy distribution of a hadron inside a jet with multiplicity $N$ equals to (see Ref.~\cite{bib:ukee})

\be
 f_N(\epsilon) = A_{mc}\; (1-x)^{D(N-1)-1}\;,
\ee{mic2}
where $x = \epsilon/E_{jet}$, $\epsilon$ is the energy of the hadron, $D$ is the effective dimensionality of the jet and $A_{mc} = {D\,N - 1 \choose D}\, DN /(\, k_D\, E_{jet}^D\,)$ follows from the normalisation condition

\be
N = \int d\Omega_p \int dp\,p^{D-1} f_N(\epsilon)\;,
\ee{mic3}
with $k_D = \int d\Omega_p$ being the angular part of the momentum space integral. Eq.~(\ref{mic2}) follows from the microcanonical momentum space volume at fixed energy and multiplicity,

\ba
\Omega_{N}(E) &=& \frac{1}{N!}\int \prod_{i=1}^{N} d^Dp_i \:\delta\left( E-\sum_{j=1}^N\epsilon_j\right)\nl
&=& \frac{\bigl[k_D\,\Gamma(D)\bigr]^N}{N!\,\Gamma(DN)} E^{N\,D\,-\,1}\;,
\ea{mic4}
and the one-particle distribution is obtained as

\be
f_N(\epsilon) = \frac{\Omega_{N-1}(E-\epsilon)}{\Omega_N(E)}\;.
\ee{mic5}
As particles in a jet form a microcanonical ensemble, an entropy $S_{jet} = \ln \Omega_{N}(E_{jet})$ and so a thermodynamical temperature

\be
\frac{1}{T_{jet}} =  \frac{\partial S_{jet}}{\partial E_{jet}} = \frac{DN-1}{ E_{jet}}
\ee{mic9a}
based on the \textit{zeroth law of thermodynamics} \cite{bib:ZerothLaw,bib:IsThereT} can be associated to them.

Microcanonical treatment of hadron production has also been proposed in Refs.~\cite{bib:ukee}~-~\cite{bib:micro7} for $e^+e^-$, $pp$ and $AA$ reactions. The main difference between our approach and the ones discussed in \cite{bib:micro2}~-~\cite{bib:micro7} is that in order to analyse the distribution of charge-averaged hadrons  inside jets of very high energy and small jetcone, we do not deal with charge conservation and neglect masses and transverse momenta of hadrons (transverse with respect to the jet direction). Thus, jet masses are neglected in our calculations: $M^2_{jet} = \left(\sum  p^\mu_i\right)^2 = \left(\sum  \epsilon_i\right)^2 - \left(\sum  \mathbf{p}_i\right)^2 \approx 0$. Consequently, the conservation of four-momentum is equivalent to energy conservation inside a one-dimensional directed jet. This way, instead of the jet mass $M_{jet}$ the jet energy $E_{jet}$ would control the distribution of hadron multiplicity. This multiplicity distribution we do not derive here, we rely on empirical fits to measurements instead.

In Refs.~\cite{bib:beck2}~-~\cite{bib:wilk2}, it has been shown that special event-by-event fluctuation patterns of the temperature or of the particle multiplicity can result in power-law tailed average particle spectra. This applies even if in each event, particles are distributed according to the Boltzmann\,--\,Gibbs distribution.
In Refs.~\cite{bib:kno1}~-~\cite{bib:kno5}, it has been argued that an approximate Koba\,--\,Nielsen\,--\,Olesen (KNO) scaling of the multiplicity distribution of charged hadrons holds for electron-positron collisions (though the scaling is weakly violated by the scale evolution of the strong fine structure constant $\alpha_s(Q^2)$).

If we consider multiplicity fluctuations of the form

\be
p(N) = \frac{\beta^{\,\alpha}}{\Gamma(\alpha)} (N-N_0)^{\alpha -1} e^{-\beta\, (N-N_0)}\;,
\ee{mic6}
and microcanonical single hadron distribution inside each jet (cf. Eq.~(\ref{mic2})) the multiplicity averaged hadron spectrum becomes

\ba
\frac{1}{\sigma}\frac{d\sigma}{d^Dp} &=& \sum_{N=N_0}^{\infty} p(N)\, f_N(\epsilon)\nl
&\approx& \dfrac{A\,  ( 1-x)^{D(N_0-1)-1} } {\left( 1 -  \dfrac{q-1}{T/E_{jet}}\ln(1-x)\right)^{1/(q-1)}} \;.
\ea{mic7}
This result can be obtained by replacing the discrete sum by an integral, and using Stirling's formula $n!\approx \sqrt{2\pi n}\,(n/e)^n$. In terms of the integration variable $\xi = N-N_0$, only the highest power is taken into account.

In Eq.~(\ref{mic7}), $N_0$ is the minimal number of hadrons that must be produced in the fragmentation process. The newly introduced parameters are: $q = 1 + 1/(\alpha +D + 1)$ and $T = E_{jet}\,\beta\,/\,[D (\alpha+D+1)]$. The parameter $q$ measures the deviation of Eq.~(\ref{mic7}) from the microcanonical distribution Eq.~(\ref{mic2}). For $e^+e^-$ data, $q>1$ holds, however, in the limit of $q\rightarrow 1$, the hadron distribution

\be
\frac{1}{\sigma}\frac{d\sigma}{dx} \rightarrow A\, x^{\,D-1} ( 1-x)^{D(\overline{N}-1)-1}  \;,
\ee{mic8}
is recovered with $\overline{N} = \alpha/\beta + N_0$ being the mean multiplicity.

Since the multiplicity fluctuates from jet to jet, so does the thermodynamical temperature introduced in Eq.~(\ref{mic9a}). The distribution of $T_{jet}$ can be obtained from Eqs.~(\ref{mic9a}) and (\ref{mic6}):

\be
p(T_{jet}) = \frac{\beta^{\,\alpha}}{\Gamma(\alpha)} \frac{D}{E_{jet} }\,\theta^2 \,(\theta-\theta_0)^{\alpha -1} \,e^{-\beta\, (\theta-\theta_0)}
\ee{mic6a}
with $\theta = E_{jet}/(D\,T_{jet})$, $\theta_0 = E_{jet}/(D\,T_{jet\,0})$ and $T_{jet\,0} = E_{jet}/[D\,N_0 - 1]$. The mean value of the thermodynamical temperature is

\be
\overline{T_{jet}} = \frac{E_{jet}}{D\,\left(\overline{N} - N_0 \right)} \, \frac{\alpha}{\alpha-1} \;+\; {\cal O}\left(1\,/\,\overline{N}^{\,2}\right)\;.
\ee{mic6b}
The $T$ parameter appearing in the multiplicity averaged hadron spectrum Eq.~(\ref{mic7}) on the other hand may be referred to as ``mean equipartition temperature''. It is proportional to the average energy per particle in a jet

\ba
\left\langle \frac{\epsilon}{N-N_0} \right\rangle_{N,\vec{p}}
&=& \sum_N p(N) \int d^Dp\,f_{N}(\epsilon) \left(\frac{\epsilon}{N-N_0} \right) \nl
&=& \frac{E_{jet}}{\overline{N}-N_0} \frac{\alpha}{\alpha-1} \nl
&=& \frac{D\,T}{1-(D+2)(q-1)}  \;,
\ea{mic9}
In the limit $q\rightarrow 1$ the mean energy per particle tends to the familiar result:

\be
\left\langle \frac{\epsilon}{N-N_0} \right\rangle_{N,\vec{p}}
\quad\rightarrow\quad  D\,T\;, \qquad \left(\text{if}\; q\rightarrow 1\right)  \;.
\ee{mic9b}
It is also worth noting that if $\overline{N}\gg N_0$ the usual equipartition formula holds for $\overline{T_{jet}}$:

\be
\left\langle \frac{\epsilon}{N-N_0} \right\rangle_{N,\vec{p}}
\quad\rightarrow\quad  D\,\overline{T_{jet}}\;, \qquad \left(\text{if}\; \overline{N}/ N_0 \rightarrow 0 \right)  \;.
\ee{mic9c}
Finally, from Eqs.~(\ref{mic6b}) and (\ref{mic9}), one can conclude that

\be
T = T_{jet}\, [1 - (D+2)(q-1)] \; +\; {\cal O}\left( 1\,/\,\overline{N}^{\,2} \right) \;.
\ee{mic9d}

\section{Analysis of Fragmentation Functions Measured in $\sqrt{s}$ = 7~TeV Proton-Proton Collisions}
\label{sec:res}
In the jet-analysis reported in Refs.~\cite{bib:atlasFFpp7TeV,bib:atlasFFpp7TeV_low}, very narrow jet-cones of $R = \sqrt{\Delta\eta^2 + \Delta\phi^2} = 0.6$ were used where $\Delta\phi$ and $\Delta\eta$ are the azimuthal angle and pseudorapidity of the hadrons relative to that of the jet. ($\eta= -\ln\tan\theta$, with $\theta$ being the polar angle.) For such a jetcone, it is reasonable to make the approximation

\be
z = \dfrac{\mathbf{p}_{_h}\, \mathbf{P}_{jet}}{|\mathbf{P}_{jet}|^2} = x\,\cos\Delta\theta \approx x \;.
\ee{mic10}
Furthermore, jets may be considered to be one-dimensional bunches of ultra-relativistic particles. This way, the four-momentum of the jet can be approximated as

\ba
P^{\mu}_{ jet} &=& (M_T \cosh y, M_T \sinh y, \mathbf{P}_{T} )\nl &\approx& (P_T \cosh \eta, P_T \sinh \eta, \mathbf{P}_{T} ) \;,
\ea{mic11}
where $M_T$ and  $\mathbf{P}_{T}$ are the transverse energy and momentum of the jet, and $y = 0.5 \ln[(E_{jet}+P_{jet\,z})/(E_{jet}-P_{jet\,z})]$. In the following, we will analyse jets mainly transverse to the beam direction ($\eta = 0$), thus, we may use $E_{jet}\approx P_{T}$. Finally, the $z$ distribution of charged hadrons takes the form

\be
\frac{1}{N_{jet}}\frac{dN}{dz} \approx \dfrac{A\, z^{\,D-1} ( 1-z)^{D(N_0-1)-1} } {\left( 1 -  \dfrac{q-1}{T^{\ast}}\ln(1-z)\right)^{1/(q-1)}} \;,
\ee{mic12}
with $T^{\ast}=T/P_{Tjet}$.

In the canonical limit, $z\ll1$ Eq.~(\ref{mic2}) tends to the Boltzmann\,--\,Gibbs distribution, and Eq.~(\ref{mic12}) approaches the q-canonical distribution

\be
\frac{1}{N_{jet}}\frac{dN}{dz}
\rightarrow A  \, \Bigg[ 1+\frac{q-1}{T^{\ast}}z \Bigg]^{-1/(q-1)}   \;.
\ee{mic13}

\begin{figure}
\begin{center}
\includegraphics[width=8cm,height=8cm]{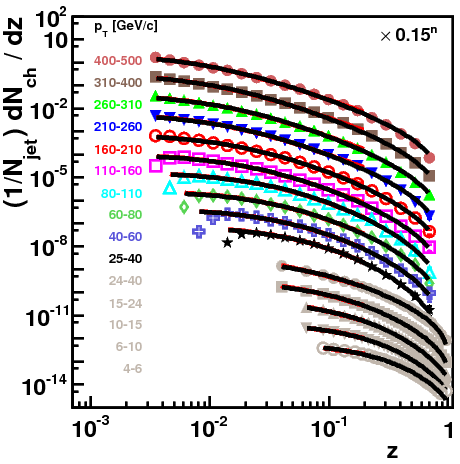} 
\end{center}
\caption{Measured distributions of the longitudinal momentum fraction $z$ of hadrons inside jets with various transverse momenta (data of jets with $P_{Tjet}= \left[4-6\right],\dots,\left[24-40\right]$ GeV/c and with $P_{Tjet}= \left[25-40\right],\dots,\left[400-500\right]$ GeV/c are published in Ref.~\cite{bib:atlasFFpp7TeV_low} and in Ref.~\cite{bib:atlasFFpp7TeV} respectively) and fitted 1 dimensional q-microcanonical distributions (Eq.~(\ref{mic12}) with $D=1$, and $N_0=3$ for the high $P_{Tjet}$ dataset and $N_0=1,1,1,2,3$ (from bottom to top) for the low $P_{Tjet}$ dataset).
\label{fig:dNdz}}
\end{figure}

\begin{figure}
\begin{center}
\includegraphics[width=8cm,height=8cm]{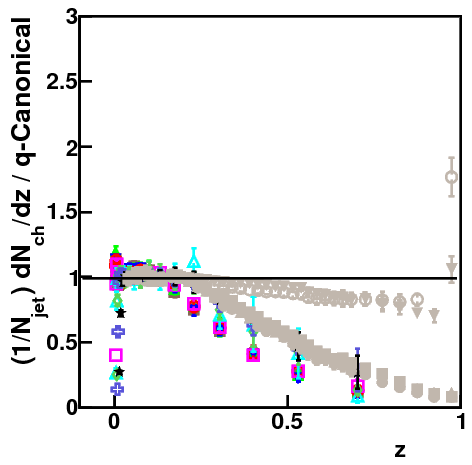} 
\end{center}
\caption{Ratios of measured $dN/dz$ distributions and fitted 1 dimensional q-canonical distributions (Eq.~(\ref{mic13}) with $D=1$) for jets with various transverse momenta (data of graphs are published in Refs.~\cite{bib:atlasFFpp7TeV_low,bib:atlasFFpp7TeV}).
\label{fig:dNdzperTS}}
\end{figure}

\begin{figure}
\begin{center}
\includegraphics[width=8cm,height=8cm]{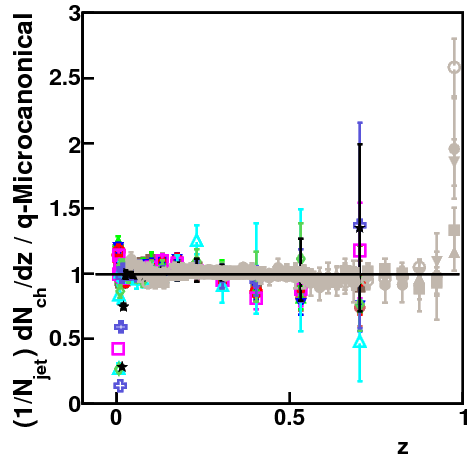} 
\end{center}
\caption{Ratios of measured $dN/dz$ distributions and fitted 1 dimensional q-microcanonical distributions (Eq.~(\ref{mic12}) with $D=1$. For the values of $N_{0}$, see the caption of Fig.~\ref{fig:dNdz}) for jets with various transverse momenta (data of graphs are published in Refs.~\cite{bib:atlasFFpp7TeV_low,bib:atlasFFpp7TeV}).\label{fig:dNdzpermTS}}
\end{figure}

\begin{figure}
\begin{center}
\includegraphics[width=8cm,height=8cm]{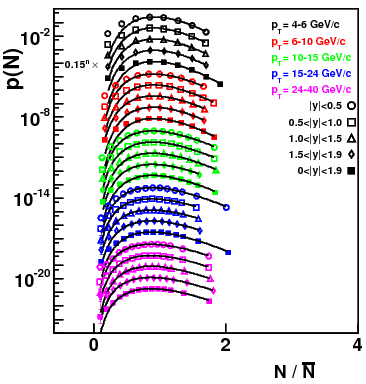} 
\end{center}
\caption{Measured multiplicity distributions of charged hadrons inside jets with various transverse momenta and rapidity and fitted Euler-Gamma distributions (Eq.~(\ref{mic6})). Data of graphs are published in Refs.~\cite{bib:atlasFFpp7TeV_low,bib:atlasFFpp7TeV}.
\label{fig:pN}}
\end{figure}

Fits of Eq.~(\ref{mic12}), (\ref{mic13}) and (\ref{mic6}) to data on fragmentation functions and multiplicity distributions measured in $pp$ collisions at $\sqrt{s}$ = 7 TeV \cite{bib:atlasFFpp7TeV,bib:atlasFFpp7TeV_low} are shown in Figs.~\ref{fig:dNdz}~-~\ref{fig:pN}. Figs.~\ref{fig:dNdz} and \ref{fig:dNdzpermTS} show that the q-microcanonical (or microcanonical Tsallis\,--\,Pareto) distribution, Eq.~(\ref{mic12}), describes data on $dN/dz$ well. The low $z$ deviation of the model from the data is assumably due to the low $p_T$ cut used in the jet analysis. Particles with transverse momentum less than $p_{T0}=0.5$ GeV/c were not taken into account in the jet analysis, and the downward curl of the measured data at low $z$ from Eq.~(\ref{mic12}) starts around $z_{0}=p_{T0}\,/P_{Tjet}$. Similar conclusions may be drawn from the data-over-theory plots shown in Fig.~\ref{fig:dNdzperTS} for the q-canonical distribution, Eq.~(\ref{mic13}), except that this distribution describes data for $z\lessapprox0.2$ only.

\begin{figure}
\begin{center}
\includegraphics[width=8cm,height=8cm]{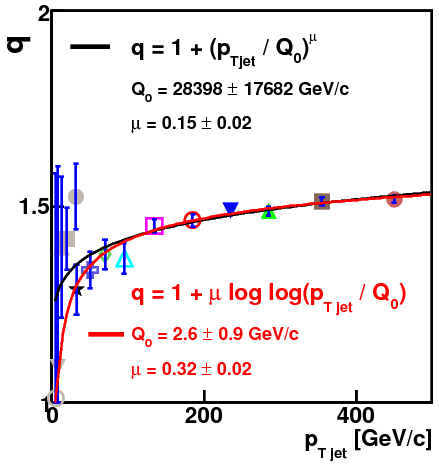} 
\end{center}
\caption{Fitted values of the $q$ parameter in Eq.~(\ref{mic13}) with $D=1$ to measured $dN/dz$ distributions shown in Fig.~\ref{fig:dNdz}.\label{fig:qTS}}
\end{figure}

\begin{figure}
\begin{center}
\includegraphics[width=8cm,height=8cm]{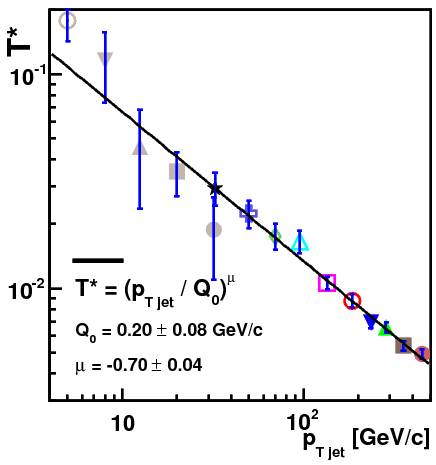} 
\end{center}
\caption{\label{fig:TTS}Fitted values of the $T^{\ast}$ parameter in Eq.~(\ref{mic13}) with $D=1$ to measured $dN/dz$ distributions shown in Fig.~\ref{fig:dNdz}.}
\end{figure}

\begin{figure}
\begin{center}
\includegraphics[width=8cm,height=8cm]{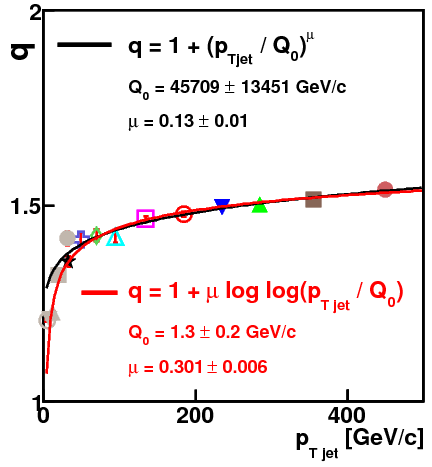} 
\end{center}
\caption{Fitted values of the $q$ parameter in Eq.~(\ref{mic12}) with $D=1$ (for the values of $N_{0}$, see the caption of Fig.~\ref{fig:dNdz}) to measured $dN/dz$ distributions shown in Fig.~\ref{fig:dNdz}.\label{fig:qMTS}}
\end{figure}

\begin{figure}
\begin{center}
\includegraphics[width=8cm,height=8cm]{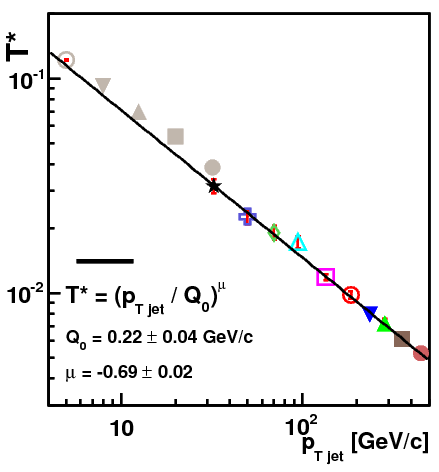} 
\end{center}
\caption{Fitted values of the $T^{\ast}$ parameter in Eq.~(\ref{mic12}) with $D=1$ (for the values of $N_{0}$, see the caption of Fig.~\ref{fig:dNdz}) to measured $dN/dz$ distributions shown in Fig.~\ref{fig:dNdz}.\label{fig:TmTS}}
\end{figure}

The evolution of the fitted $q$ and $T$ parameters of Eqs.~(\ref{mic12}) and (\ref{mic13}) with the transverse momentum of the jet are shown in Figs.~\ref{fig:qTS}~-~\ref{fig:TmTS}. The ``mean equipartition temperature'' parameter scaled by the transverse momentum of the jet $T^{\ast} = T/P_{Tjet}$ shows power-law dependence on $P_{Tjet}$,

\be
T^{\ast} = \left(P_{Tjet}/Q_{0} \right)^{\mu}\;,
\ee{mic14}
while for the parameter $q$, both a power-law,

\be
q = \left(P_{Tjet}/Q_{0} \right)^{\mu}\;,
\ee{mic15}
and a double-logarithmic ansatz,

\be
q = 1 + \mu \ln\ln \left(P_{Tjet}/Q_{0} \right)\;,
\ee{mic16}
fit. The $Q_0$ and $\mu$ parameters of the q-canonical and the q-microcanonical distributions approximately coincide within errors.

For $q$, Eq.~(\ref{mic16}) was successfully used in Ref.~\cite{bib:Gribov80} to adjust the $Q^2$ evolution of a fragmentation function of the form of Eq.~(\ref{mic13}) to that of an AKK type one \cite{bib:DzAKK}. In $pp$ collisions, $Q=P_{Tjet}$ seems to be a good choice. The $q$ and $T$ parameters show similar scale dependence for fragmentation functions of protons, $K^{0}$-s, $\pi^{0}$-s, $\Lambda$-s and charge-averaged hadrons produced in $e^+e^-$ annihilations as well as for the transverse momentum spectra of charged hadrons stemming from $pp$ collisions  \cite{bib:ukee,bib:ukeeL,bib:ukeeK,bib:HotQuarks2010}.

Fig.~\ref{fig:pN} shows that Eq.~(\ref{mic6}) describes data on multiplicity distributions well, except for $N<3$, where measured data are higher than what the Euler-Gamma distribution predicts. This effect is perhaps due to the small conesize. In a jet with one or two very energetic particles, the others have very small energies, and thus may fly out of the jetcone. This way, the number of jets with only a few particles increases in this type of jet analysis. As a consequence, the $\alpha$ parameter of Eq.~(\ref{mic6}), which is greatly influenced by the number of low multiplicity jets, can not be determined reliably. Thus, it is not so disconcerting that the $q$ parameter predicted from multiplicity fits takes lower values (of around $q=1.1$) than that obtained from fits to $dN/dz$ data.

It is important to note that in Ref.~\cite{bib:atlasFFpp7TeV_low}, jets were reconstructed from charged particles only, while in Ref.~\cite{bib:atlasFFpp7TeV}, calorimetric measurements were used in the jet reconstruction, so
both neutral and charged particles were included in the analysis. For this reason, we fitted Eqs.~(\ref{mic14})~-~(\ref{mic16}) to $q$ and $T$ values obtained for jets with high transverse momenta only (25 GeV/c $\leq P_{Tjet} \leq$ 500 GeV/c from \cite{bib:atlasFFpp7TeV}). Nevertheless, $q$ and $T$ values for jets with low transverse momenta (4 GeV/c $\leq P_{Tjet} \leq$ 40 GeV/c obtained from \cite{bib:atlasFFpp7TeV_low}) show a tendency similar to that of the high $P_{Tjet}$ jets. The $T^{\ast}$ parameters of the low $P_{Tjet}$ dataset are approximately 10\% higher, than what the fit of Eq.~(\ref{mic14}) to the high $P_{Tjet}$ dataset predicts (see Fig.~\ref{fig:TmTS}). From Eq.~(\ref{mic9}), it can be seen that $T^{\ast}$ is proportional to the inverse of the multiplicity. This way, the ratio of $T^{\ast}$-s obtained in the two different analysis' is proportional to the ratio of the total multiplicity to that of charged particles. If as an estimate, we used the ratio of charged to neutral pions, we would get a factor of 3/2 for the ratio of $T^{\ast}$-s obtained from the two different analysis'. This value is somewhat higher than what can be seen in Fig.~\ref{fig:TmTS}.

Eq.~(\ref{mic12}) describes both datasets. For the high $P_{Tjet}$ dataset, the power of the $1-z$ factor in the numerator takes the value $D(N_0-1)-1 = +1$, while this quantity decreases from 1 to -1 for the low $P_{Tjet}$ dataset as $P_{Tjet}$ decreases from 40 GeV/c to 4 GeV/c.

\section{Conclusions}
\label{sec:con}
This paper shows that the statistical jet-fragmentation model \cite{bib:ukee,bib:ukeeL} describes the $dN/dz$ distribution of hadrons in jets created in proton-proton reactions at $\sqrt{s}$ = 7 TeV center of mass energy \cite{bib:atlasFFpp7TeV,bib:atlasFFpp7TeV_low}. This model combines microcanonical statistics (which has also been used in the description of different hadronic observables in high-energy phenomena in Refs.~\cite{bib:micro1}-\cite{bib:micro6}) with super-statistics \cite{bib:beck2}-\cite{bib:wilk2} stemming from multiplicity fluctuations emerging in proton-proton as well as in electron-positron and nucleus-nucleus collisions \cite{bib:kno1}-\cite{bib:kno5}.

It turns out that the parameters of the $dN/dz$ distribution of charged hadrons in jets in proton-proton collisions (Sect.~\ref{sec:res}) show similar scale dependence as the parameters of fragmentation functions in electron-positron annihilations \cite{bib:ukee,bib:ukeeL,bib:ukeeK} and of transverse momentum spectra of charged hadrons stemming from proton-proton collisions \cite{bib:HotQuarks2010}. These scale evolutions are consistent with the DGLAP equations \cite{bib:Gribov80}.

Finally, it is pointed out that the $dN/dz$ distributions obtained from two different jet analysis' and different kinematical ranges \cite{bib:atlasFFpp7TeV,bib:atlasFFpp7TeV_low} both can be described by the microcanonical jet-fragmentation model.

\section*{Acknowledgement}
\label{sec:ack}
The authors thank Brian Cole and Eric Feng for the discussions. This work was supported by the Hungarian OTKA grants K68108, K104260, NK77816 and by the bilateral Hungarian--South-African project NIH TET 10-1 2011-0061, ZA-15/2009. One of the authors (GGB) thanks the J\'anos Bolyai Research Scolarship from the Hungarian Academy of Sciences.





\begin{thebibliography}{00}




\bibitem{bib:pQCD}
Y. Zhang et al, \textit{Phys. Rev. C}, \textbf{65}, 034903 (2002).

\bibitem{bib:DzAKK}
S. Albino, B. A. Kniehl, G. Kramer, \textit{Nucl. Phys. B}, \textbf{725}, 181 (2005).

\bibitem{bib:DzKretzer}
S. Kretzer, \textit{Phys. Rev. D}, \textbf{62}, 054001 (2000).






\bibitem{bib:DGLAPgl}
V. N. Gribov, L. N. Lipatov, \textit{Sov. J. Nucl. Phys.}, \textbf{15} 438 (1972).

\bibitem{bib:DGLAPap}
G. Altarelli, G. Parisi, \textit{Nucl. Phys. B} \textbf{126} 298 (1977)

\bibitem{bib:DGLAPd}
Yu. L. Dokshitzer, \textit{Sov. Phys. JETP}, \textbf{46} 641 (1977).


\bibitem{bib:ukee}
K. Urmossy, G. G. Barnaf\"oldi, T. S. Bir\'o, \textit{Phys. Lett. B}, \textbf{701}, 111-116, (2011)

\bibitem{bib:ukeeL}
K. Urmossy, et. al., \textit{Acta Physica Polonica B}, \textbf{5}:(2), pp. 363-368, (2012) 

\bibitem{bib:ukeeK}
T. S. Biro, et. al., \textit{Acta Physica Polonica B}, \textbf{43}:(4), pp. 811-820, (2012) 

\bibitem{bib:micro1}
V. V. Begun, M. Gazdzicki, M. I. Gorenstein, \textit{Phys. Rev. C:} \textbf{78}, 024904, (2008)

\bibitem{bib:micro1.1}
M. Hauer, V. V. Begun, M. Gazdzicki, M. I. Gorenstein, V. P. Konchakovski, B. Lungwitz, \textit{J. Phys. G} \textbf{35} 044064, (2008) 

\bibitem{bib:micro1.2}
V. V. Begun, M. Gazdzicki, M. I. Gorenstein, \textit{Acta Physica Polonica B}, \textbf{43} 1713, (2012) 

\bibitem{bib:micro2}
F. M. Liu, K. Werner, J. Aichelin, \textit{Phys. Rev. C:} \textbf{68}, 024905, (2003) 

\bibitem{bib:micro3}
F. M. Liu, K. Werner, J. Aichelin, M. Bleicher, H. Stoecker, \textit{J. Phys. G:} \textbf{30}, 589-594, (2004)

\bibitem{bib:micro4}
F. M. Liu, K. Werner, \textit{Phys. Rev. D:} \textbf{74}, 034024, (2006)

\bibitem{bib:micro5}
F. Becattini, L. Ferroni, \textit{Eur. Phys. J. C:} \textbf{35}, 243-258, (2004)

\bibitem{bib:micro6}
F. Becattini, L. Ferroni, \textit{Eur. Phys. J. C:} \textbf{38}, 225-246 (2004)

\bibitem{bib:micro7}
C. Bignamini, F. Becattini, F. Piccinini, arXiv:1204.2300v1 (2012)








\bibitem{bib:kno1}
A. M. Polyakov, \textit{Zh. Eksp. Teor. Fiz.} \textbf{59}, 542 (1970)

\bibitem{bib:kno2}
Z. Koba, H. B. Nielsen, P. Olesen, \textit{Nucl. Phys. B} \textbf{40}, 317 (1972)

\bibitem{bib:kno3}
S. Hegyi, \textit{Phys. Lett. B:} \textbf{467}, 126-131, (1999)

\bibitem{bib:kno4}
S. Hegyi, \textit{Proc. ISMD 2000, Tihany, Lake Balaton, Hungary}, (2000)

\bibitem{bib:kno5}
Yu. L. Dokshitzer, \textit{Phys. Lett. B}, \textbf{305}, 295 (1993); LU-TP/93-3 (1993)


\bibitem{bib:beck2}
C. Beck, E. G. D. Cohen, \textit{Physica A} \textbf{322}, 267, (2003)

\bibitem{bib:beck3}
C. Beck, E. G. D. Cohen, \textit{Physica A} \textbf{344}, 393, (2004)

\bibitem{bib:beck4}
H. Touchette, C. Beck, \textit{Phys. Rev. E} \textbf{71}, 016131, (2005)

\bibitem{bib:beck5}
S. Abe, C. Beck, G. D. Cohen, \textit{Phys. Rev. E} \textbf{76}, 031102, (2007)

\bibitem{bib:wilk1}
M. Biyajima, T. Mizoguchi, N. Nakajima, N. Suzuki, G. Wilk, \textit{Eur. Phys. J. C:} \textbf{48}, 597-603, (2006) 

\bibitem{bib:wilk2}
G. Wilk, Z. Wlodarczyk \textit{Eur. Phys. J A} \textbf{40}, 299, (2009)


\bibitem{bib:ZerothLaw}
T. S. Biro, P. V\'an, \textit{Phys. Rev. E}, \textbf{83} 061147, (2011) 

\bibitem{bib:IsThereT}
T. S. Biro, \textit{Is there a temperature?}, Springer, 2011. (ISBN 978-1-4419-8040-3), (2011)


\bibitem{bib:atlasFFpp7TeV}
G. Aad et al. [ATLAS Collaboration], \textit{Eur. Phys. J. C}, \textbf{71} 1795 (2011)

\bibitem{bib:atlasFFpp7TeV_low}
G. Aad et al. [ATLAS Collaboration], \textit{Phys. Rev. D}, \textbf{84} 054001 (2011) 



\bibitem{bib:HotQuarks2010}
G. G. Barnaf\"oldi et al., \textit{J. Phys. Conf. Ser.}, \textbf{270}, 012008, (2011) 

\bibitem{bib:Gribov80}
G. G. Barnaf\"oldi et. al., \textit{Gribov 80 - Memorial Volume}, Singapore, World Scientific, 2011. p. 357. (ISBN: 978-981-4350-18-1), (2011) 
\end{thebibliography}
\end{document}